\documentclass[a4paper,fleqn,usenatbib]{mnras}

\usepackage{newtxtext,newtxmath}
\usepackage[T1]{fontenc}
\usepackage{ae,aecompl}

\usepackage{graphicx}	% Including figure files
\usepackage{amsmath}	% Advanced maths commands
\usepackage{amssymb}	% Extra maths symbols

\title[Interstellar dimethyl ether gas-phase formation]{Interstellar dimethyl ether gas-phase formation: a quantum chemistry and kinetics study}

\author[D. Skouteris et al.]
{Dimitrios Skouteris$^{1}$\thanks{Present address: Master-Up, Via Elce di Sotto 8, I-06123 Perugia, Italy},
Nadia Balucani$^{2,3,4}$\thanks{e-mail:nadia.balucani@unipg.it},
Cecilia Ceccarelli$^{3}$,
Noelia Faginas Lago$^{2}$,
\newauthor
Claudio Codella$^{4,3}$,
Stefano Falcinelli$^{5}$,  
Marzio Rosi$^{5}$
\\
\\
$^1$Scuola Normale Superiore, Piazza dei Cavalieri 7, I-56126 Pisa, Italy\\
$^2$Dipartimento di Chimica, Biologia e Biotecnologie, Universit\`a di Perugia, Via Elce di Sotto 8, I-06123 Perugia, Italy\\
$^3$Institut de Plan\'etologie et d'Astrophysique de Grenoble (IPAG), rue de la Piscine, F-38041, Grenoble, France\\
$^4$INAF-Osservatorio Astrofisico di Arcetri, largo E. Fermi 5, I-50125, Firenze, Italy\\
$^5$Dipartimento di Ingegneria Civile ed Ambientale, Via Duranti 93, I-06125 Perugia, Italy
}
\date{Accepted XXX. Received YYY; in original form ZZZ}

\pubyear{2018}

\begin{document}
\label{firstpage}
\pagerange{\pageref{firstpage}--\pageref{lastpage}}
\maketitle

\begin{abstract} 
Dimethyl ether is one of the most abundant interstellar complex organic molecules. Yet its formation route remains elusive. In this work, we have performed electronic structure and kinetics calculations to derive the rate coefficients for two ion-molecule reactions recently proposed as a gas-phase formation route of dimethyl ether in interstellar objects, namely CH$_3$OH + CH$_3$OH$_2^+$~$\rightarrow$~ (CH$_3$)$_2$OH$^+$ + H$_2$O followed by (CH$_3$)$_2$OH$^+$ + NH$_3$~ $\rightarrow$~ CH$_3$OCH$_3$ + NH$_4^+$. A comparison with previous experimental rate coefficients for the reaction CH$_3$OH + CH$_3$OH$_2^+$ sustains the accuracy of the present calculations and allow a more reliable extrapolation at the low temperatures of interest in interstellar objects (10-100 K). The rate coefficient for the reaction (CH$_3$)$_2$OH$^+$ + NH$_3$ is, instead, provided for the first time ever.  The rate coefficients derived in this work essentially confirm the prediction by Taquet et al. (2016) concerning dimethyl ether formation in hot cores/corinos. Nevertheless, this formation route cannot be efficient in cold objects  (like prestellar cores) where dimethyl ether is also detected, because ammonia has a very low abundance in those environments.
\end{abstract}


%%%%%%%%%%%%%%%%%%%%%%%%%%%%%%%%%%%%%%%%%%%%%%%%%%

%%%%%%%%%%%%%%%%% BODY OF PAPER %%%%%%%%%%%%%%%%%%

% The MNRAS class isn't designed to include a table of contents, but for this document one is useful.
% I therefore have to do some kludging to make it work without masses of blank space.
%\begingroup
%\let\clearpage\relax
%\tableofcontents
%\endgroup

\section{Introduction}\label{sec:Intro}
Since their first detection in the interstellar medium (ISM), the presence of relatively complex organic molecules (from now on indicated with iCOMs for interstellar Complex Organic Molecules, namely C-bearing species with at least six atoms: e.g. Ceccarelli et al. 2017; Herbst \& van Dishoeck 2009) has posed the question of how they are formed. The harsh chemical environments of interstellar clouds (namely, very low temperature and very low number density), indeed, challenge the common notions that chemical synthesis requires energy to promote the weakening of the reactant bonds and frequent collisions to increase the number of reactive encounters. Since about 1\% of interstellar clouds is composed by submicron sized silicates and vitreous graphite particles, interstellar grains covered by icy mantles are also invoked to play an important role in synthesizing iCOMs by acting as interstellar catalysts (e.g. Garrod \& Herbst 2006; Taquet et al. 2012; Ag\`undez \& Wakelam 2013). Recent astrochemical models are able to include both gas-phase processes and grain-induced chemistry in an attempt to reproduce the observed iCOMs abundances (e.g. Garrod et al. 2008; Balucani et al. 2015; Skouteris et al. 2017, 2018; Ruaud et al. 2016; Taquet et al. 2016; Vasyunin  et al. 2017; Quenard et al. 2018). Still, problems remain in accounting for all observed species in different interstellar objects.

One important drawback of all models, which include several thousands of molecular processes, is due to the uncertainty associated to the parameters which are used to quantitatively account for the importance of every step. Many of those processes have never been investigated in laboratory experiments, many others have been investigated but under experimental conditions that do not reproduce the interstellar ones (either regarding the temperature or the pressure or UV illumination). For gas-phase reactions of the first kind, rate coefficients and their temperature dependence are mainly estimated with some chemical intuition or by drawing analogies with similar known processes. Small details in the molecular structure, however, can induce a huge change in the chemical behaviour and reasoning by analogy can cause severe mistakes. In the second case, the values obtained as a function of the temperature in a temperature range that does not encompass those of relevance in ISM are used, but this can also be very risky as a change in the reaction mechanism can alter the temperature dependence in non-Arrhenius reactions. In this respect, recent kinetics experiments performed with the CRESU ({\it Cin\'etique de R\'eaction en Ecoulement Supersonique Uniforme}, in English Reaction Kinetics in Uniform Supersonic Flow) technique have shown that the reactions characterized by a pre-reactive complex with some stability can be characterized by a very large rate coefficient at low temperatures even though their values at room temperature are affected by the presence of an energy barrier (for a recent review on these cases, see Potapov et al. 2017; see also Georgievskii \& Klippenstein 2007). The case of grain-chemistry simulations in laboratory experiments is even more complex, as no experiments are able to reproduce the size of interstellar particles, the exact composition of the grain icy mantle and the flux of particles and/or photons impinging on the grains (e.g. Linnartz et al. 2015).

A theoretical characterization at the atomic/molecular level can help in extrapolating experimental data at the conditions of ISM or in estimating in a reliable way the kinetic parameters associated to reactions that cannot be investigated in laboratory experiments. For this reason, several of the authors of this paper have started a systematic investigation of gas-phase bimolecular reactions involving either neutral or charged species for which no data (Balucani et al. 2015; Skouteris et al. 2015; Barone et al 2015; Vazart et al. 2015; Skouteris et al. 2017, 2018; Rosi et al. 2018) or data limited at high temperature or pressure conditions (Balucani et al. 2012; Leonori et al. 2013; Balucani et al. 2015; Sleiman et al. 2018) are available. On the same vein, several studies have been carried out to try to investigate, at an atomic level, reactions occurring on the iced surfaces of the interstellar grains (e.g. Rimola et al. 2014, 2018; Enrique-Romero et al. 2016; Song \& Kastner 2016, 2017; Lamberts 2018).
The goal of all these studies is a better understanding of the involved processes and the increase of the accuracy of the parameters employed in astrochemical models. This will have, hopefully, the consequence of having models with an improved capability of predicting the observed abundances of iCOMs. Successful examples include formamide (Barone et al. 2015; Skouteris et al. 2017; Codella et al. 2017) and glycolaldehyde (Skouteris et al. 2018).

In this contribution, we present a theoretical characterization of the two-reaction sequence which has been suggested by Charnley \& co-workers (Charnley et al. 1995; Rodgers \& Charnley 2001; Taquet et al. 2016) to produce dimethyl ether (one of the most abundant and ubiquitous iCOMs) in the gaseous phase, namely:
\begin{tabbing}
  CH$_3$OH + CH$_3$OH$_2^+$ \= $\rightarrow$ \= (CH$_3$)$_2$OH$^+$ + H$_2$O ~~~~~~\=  (1)\\
(CH$_3$)$_2$OH$^+$ + NH$_3$ \> $\rightarrow$ \> CH$_3$OCH$_3$ + NH$_4^+$ \> (2)
\end{tabbing}
Reaction (1) is present in the two major databases repository of astrochemistry reactions used by different modellers: KIDA ({\it kida.obs.u-bordeaux1.fr}: Wakelam et al. 2012) and UMIST ({\it http://udfa.ajmarkwick.net}: McElroy et al. 2013).  On the contrary, reaction (2) is not present in either the UMIST or KIDA database. The (potential) efficacy of this reaction sequence has been demonstrated in a recent paper by Taquet et al. (2016). In particular, the proton transfer to ammonia, for this and other protonated iCOMs, seems to be able to compensate for the missing role of electron-ion recombination processes. Those processes, which were supposed to convert the molecular ions easily built via ion-molecule reactions into their neutral counterparts (the actually observed species), proved to be mainly of dissociative kind for iCOMs some years ago (e.g. Geppert et al. 2005). This is indeed the case of protonated dimethyl ether, as demonstrated by Hamberg et al. (2010): only 7\% of protonated dimethyl ether (in its perdeuterated isotopologue, (CD$_3$)$_2$OD$^+$) was experimentally determined to eject a single hydrogen (D) atom, while 49\% of the reaction outcome is associated with the break-up of the C--O--C chain and 44\% with the rupture of both C--O bonds. 

To verify whether the reactions (1)-(2) can play the role suggested by Taquet et al. (2016), we have carried out dedicated electronic structure calculations of the two relevant potential energy surfaces (PESs) and kinetics calculations to derive rate coefficients as a function of the temperature under collision-free conditions, as those characterizing ISM gas. This manuscript is organized as follows. In Section \ref{sec:past-studies-two}, we briefly summarise what is known for reactions (1) and (2).  In Section \ref{sec:computations}, the employed theoretical methods are described as well as the results of electronic structure and kinetics calculations. Discussion and astrochemical implications are presented in Section \ref{sec:discussion}.

% % % %
% % % %
% % % % 
%%%%%%%%%%%%%%%%%%%%%%%%%%%%%%%%%%%%%%%%%%%%%%%%%%%%%%%
\section{Previous studies on the two reactions}\label{sec:past-studies-two}
\subsection{The reaction CH$_3$OH + CH$_3$OH$_2^+$}\label{sec:reaction-ch_3oh-+}

There are numerous experimental investigations by means of various experimental techniques exploring different pressure (from 1-10$^{-4}$ mbar in ion flow tube experiments to 10$^{-6}$--10$^{-7}$ mbar in ion cyclotron resonance experiments) and temperature (from 293 K to 670 K) ranges (Karpas \& Meot-Ner 1989; Morris et al. 1991; Dang \& Bierbaum 1992; Fridgen et al. 2001). The results are in partial disagreement, especially on the product branching ratio. There are two possible outcomes for this process in addition to the methyl transfer reaction (1), that is, proton transfer (3) and adduct formation (4) as listed below:
\begin{tabbing}
  CH$_3$OH + CH$_3$OH$_2^+$ \= $\rightarrow$ \= CH$_3$OH$^+$ + CH$_3$OH ~~~~~~\=  (3)\\
  CH$_3$OH + CH$_3$OH$_2^+$ \= $\rightarrow$ \= (CH$_3$OH)$_2$H$^+$\> (4)
\end{tabbing}
Different experiments, performed with different techniques and, especially, under different pressure conditions, have provided different branching ratios for channels (1),(3) and (4). Nevertheless, at room temperature the absolute value for the rate coefficient of reaction (1) falls, in all cases, in the range 0.8-1.0 $\times10^{-10}$ cm$^3$ molec$^{-1}$ s$^{-1}$. No experimental data are available at the very low temperatures of interest in interstellar objects. In addition to experimental studies, a first theoretical characterization of the reaction mechanism has been performed by Bouchoux \& Choret (1997) at the MP2/6-31G*//MP2/6-31G* + ZPE level of calculations. The energy of several stationary points has also been calculated more recently by Fridgen et al. (2001) at the MP2/6-311G** level and basis set to assist the interpretation of their experimental findings. Nonetheless, in none of the previous theoretical investigations was a kinetics analysis attempted. Interestingly, by referring to the experimental results on isotopically labelled reactions (with D and $^{18}$O), Bouchoux \& Choret (1997) and Fridgen et al. (2001) reached opposite conclusions on the initial approach of the reactants: according to the former authors, the formation of a hydrogen bond adduct is dominant while according to the latter, the yield of isotopically labelled products indicates that the first step is an S$_N$2 attack. In addition, the experimental value of the activation energy ($-49.8\pm1.7)$ kJ/mol) is in excellent agreement with the theoretical one for the S$_N$2 attack and in great disagreement with that associated to the hydrogen-bond adduct.

\begin{figure}
  \centering
  \includegraphics[width=8.0cm]{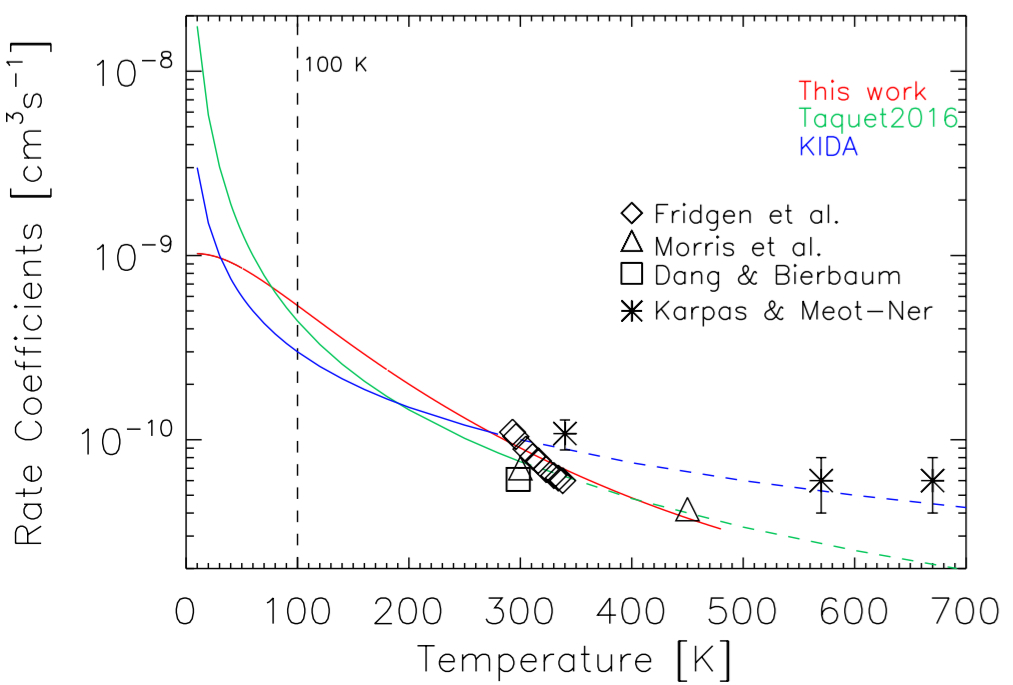}
  \caption{Rate of the reaction CH$_3$OH + CH$_3$OH$_2^+$~$\rightarrow$~ (CH$_3$)$_2$OH$^+$ + H$_2$O, as reported in the KIDA and UMIST (blue) databases (see text), Taquet et al. (2016) (green) and as computed this work (red). Dashed lines report the KIDA and Taquet et al. (2016) extrapolated rates in the 300--700 K range. The values previously computed or obtained in laboratory experiments are reported with different symbols (see text): Fridgen et al. (2001), Morris et al. (1991), Dang \& Bierbaum 1992), Karpas \& Meot-Ner (1989). The horizontal dashed black line indicates the temperature used in Taquet et al. (2016) modelling.}
  \label{fig:rates}
\end{figure}

The temperature dependence of the rate coefficient has been experimentally determined in a very limited range of temperatures by Karpas \& Meot-Ner (1989), Morris et al. (1991), and by Fridgen et al. (2001). In all cases, the explored temperatures are far from the range of interest in interstellar chemistry.  The available experimental rate coefficient as a function of the temperature are summarized in Fig. \ref{fig:rates}, whete also the KIDA and UMIST values and their trend with T are shown (in a recent update, the value adopted in UMIST is the same as that adopted by KIDA). In their model, Taquet et al. (2016) used the old UMIST values, with $\alpha$ = $7.6\times10^{-11}$ cm$^3$s$^{-1}$ and $\beta$ = -1.6.
We recall, however, that the experimental data of Fig. 1 are all obtained in a very different T range (between 293 and 670 K) and that the extrapolation of the T dependence outside the range of explored temperature is not warranted.

In this respect, we note that KIDA recommendation relies on the extrapolation of the T dependence determined by Karpas \& Meot-Ner (1989), while the old UMIST recommendation used by Taquet et al. (2016) relies on the extrapolation of the T dependence determined by Morris et al. (1991).

\subsection{The reaction (CH$_3$)$_2$OH$^+$ + NH$_3$}
To the best of our knowledge, there are no experimental data on this process or previous theoretical investigations. In their network of reactions, Taquet et al. (2016) have employed a value of the rate coefficient of $2\times10^{-9}$ cm$^3$s$^{-1}$ for all the proton transfer reactions involving ammonia and protonated iCOMs. This choice is given by the fact that the experimental values derived by Hemsworth et al. (1974) for a series of proton transfer reactions involving ammonia are all very similar and in the range ($2\pm1$) $\times 10^{-9}$ cm$^3$s$^{-1}$. Nevertheless, we would like to mention that the proton affinity of dimethyl ether is higher than those associated to the species investigated by Hemsworth et al. (1974), being 792 kJ/mol as opposed to 422 for H$_2$ (the lowest) and 751 kJ/mol for C$_3$H$_6$ (the highest). In other words, the difference between the proton affinities of ammonia (853 kJ/mol) and dimethyl ether (which corresponds to the enthalpy variation associated to the proton transfer process) is rather smaller than those associated to most of the reactions characterised by Hemsworth et al. (1974). According to the Hammond's postulate, therefore, the rate coefficient for the process (2) should be in the lower limit of the values recorded by Hemsworth et al. (1974), that is ca. 1$\times 10^{-9}$ cm$^3$ molec$^{-1}$ s$^{-1}$.

% % % %
% % % %
% % % % 
%%%%%%%%%%%%%%%%%%%%%%%%%%%%%%%%%%%%%%%%%%%%%%%%%%%%%%%
\section{COMPUTATIONAL METHODS AND RESULTS}\label{sec:computations}
In this Section, we first provide details on the method employed to obtain the stationary points of the PESs of the two studied reactions, followed by the results of these calculations. Then, we will describe the  kinetics calculations and results for the two reactions.
\begin{figure*}
  \centering
  \includegraphics[width=15.0cm]{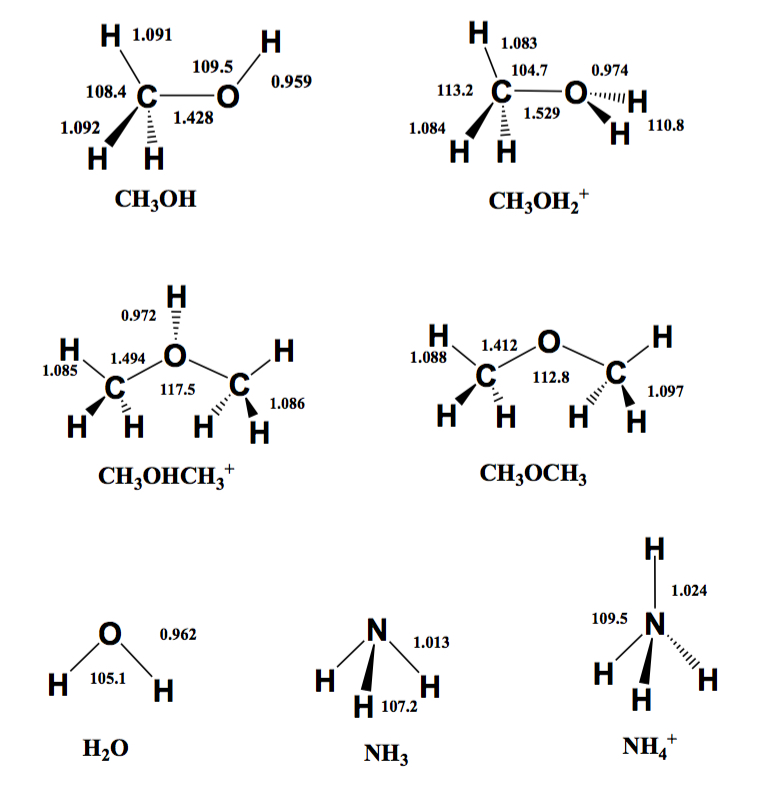}
  \caption{B3LYP optimized geometries (\AA ~and $^o$) of reactants and products of reactions (1) and (2).}
  \label{fig:geometry}
\end{figure*}

\subsection{Electronic Structure Calculations: Methods}\label{sec:deta-potent-energy}

We characterized the PES of the two reactive systems through optimization of the most stable stationary points. For this we performed density functional (DFT) calculations with the Becke 3-parameter exchange and Lee-Yang-Parr correlation (B3LYP) (Becke 1993; Stephens et al. 1994) hybrid functional, as well as the correlation consistent valence polarized set aug-cc-pVTZ (Dunning 1989; Kendall et al. 1992; Woon \& Dunning 1993). Using the same level of theory we have calculated the harmonic vibrational frequencies to determine the nature of each stationary point, i.e. minimum if there are no imaginary frequencies and saddle point if exactly one frequency is imaginary. We have assigned the saddle points through intrinsic reaction coordinate (IRC) calculations (Gonzalez \& Schlegel 1989; Gonzalez \& Schlegel 1990). Then, we computed the energy of each stationary point with the more accurate coupled cluster theory including single and double excitations as well as a perturbative estimate of connected triples (CCSD(T)) (Bartlett 1981; Raghavachari et al. 1989; Olsenet et al. 1996) using the same basis set aug-cc-pVTZ. We have added the zero point energy correction to both energies (calculated by B3LYP and CCSD(T)) to correct them to 0 K. This correction was computed using the scaled harmonic vibrational frequencies obtained at the B3LYP/aug-cc-pVTZ level. All calculations were carried out using Gaussian 09 (Frisch et al. 2009) and the vibrational analysis was performed using Molekel (Flukiger et al. 2000; Portmann and Lüthi 2000).

\begin{figure*}
  \centering
  \includegraphics[width=15.0cm]{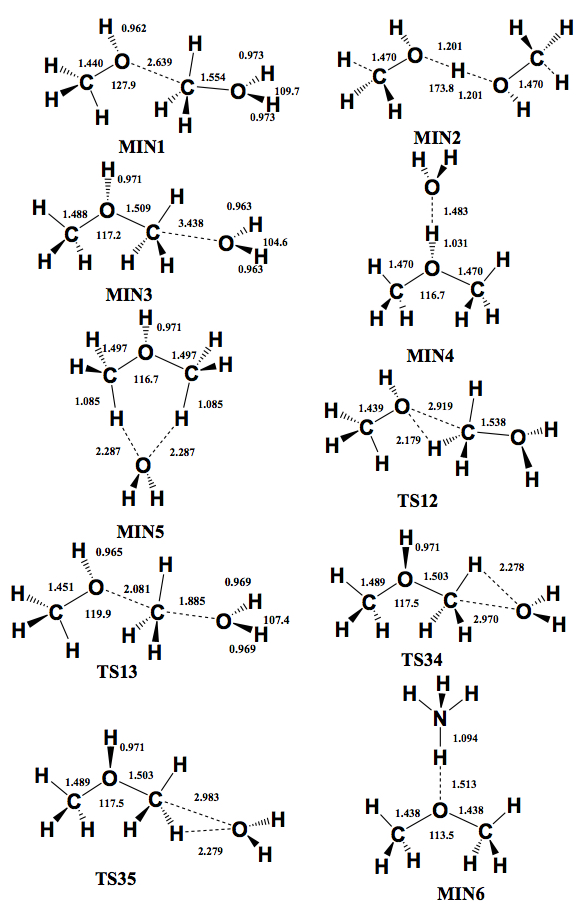}
  \caption{B3LYP optimized geometries (\AA ~and $^o$) of the investigated stationary points of the reactions 1 and 2.}
  \label{fig:reac1-geometry}
\end{figure*}

\subsection{Electronic Structure Calculations: Results}\label{sec:results-potent-energ}

\subsubsection{The reaction CH$_3$OH + CH$_3$OH$_2^+$}
\begin{figure*}
  \centering
  \includegraphics[width=17.0cm]{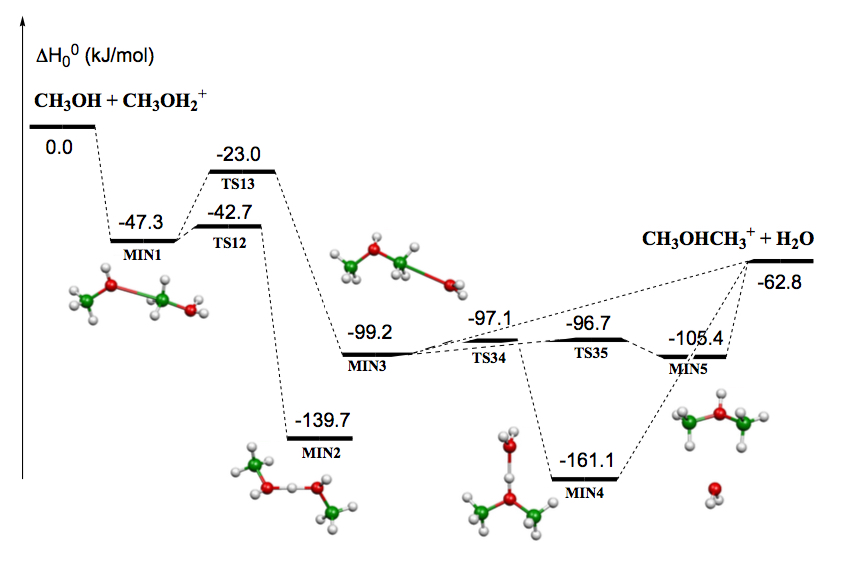}
  \caption{Schematic representation of the potential energy surface relative to the interaction between CH$_3$OH and CH$_3$OH$_2^+$. For simplicity, only the CCSD(T) relative energies (kJ/mol) are reported. }
  \label{fig:reac1-pes}
\end{figure*}
The B3LYP/aug-cc-pVTZ optimized structures of minima and saddle points involved in reaction (1) are shown in Fig. \ref{fig:geometry} and \ref{fig:reac1-geometry}, while enthalpy changes and barrier heights for each step, computed both at B3LYP7aug-cc-pVTZ and CCSD(T)/aug-cc-pVTZ levels, are reported in Table \ref{tab:enthalpy}. 
In the PES of the system [(CH$_3$OH)$_2$H]$^+$ (see Fig. \ref{fig:reac1-pes}) we localized five minima, MIN1, MIN2, MIN3, MIN4, MIN5 which are connected by four transition state, TS12 (which connects MIN1 with MIN2), TS13 (which connects MIN1 with MIN3), TS34 (which connects MIN3 with MIN4), TS35 (which connects MIN3 with MIN5). Some of these stationary points have been previously investigated. In particular, Bouchoux \& Choret (1997) characterized MIN1, MIN2, MIN3,MIN4, TS12, TS13 and TS34 at MP2/6-31G* level, while Fridgen et al.(2001) studied MIN1, MIN2, MIN3, TS12 and TS13 at MP2/6-311G** level. The agreement of our calculations with these previously reported lower level calculations is reasonable, if one considers the different methods and smaller basis sets employed. The reaction between methanol and protonated methanol starts with a barrierless formation of the adduct MIN1 characterised by a new C--O interaction, as we can see from the schematic PES reported in Fig. \ref{fig:reac1-pes}:  

CH$_3$OH + CH$_3$OH$_2^+$~~ $\rightarrow$ ~~ CH$_3$OH$\cdots$CH$_3$OH$_2^+$ ~~~~~~(a)\\

In the following discussion, for simplicity, only the more accurate CCSD(T)/aug-cc-pVTZ energies will be referred to (for the B3LYP/aug-cc-pVTZ energy values, see Table \ref{tab:enthalpy}). The process (a) leading to MIN1 is exothermic by 47.3 kJ/mol. The C--O bond length, 2.639 \AA, is very long suggesting that this could be considered as an electrostatic interaction rather than a true chemical bond. Species MIN1 through the transition state TS13, which lies under the energy of the reactants, can isomerize to species MIN3 where the C--O interaction becomes a true chemical bond (bond length 1.509 \AA) while the terminal C--O bond becomes an electrostatic interaction, as we can notice from the C--O bond length which changes from 1.554 \AA~ to 3.438 \AA. The step

\noindent
CH$_3$OH$\cdots$CH$_3$OH$_2^+$(MIN1)~$\rightarrow$~CH$_3$OHCH$_3$(MIN3)
$\cdots$OH$_2^+$ ~~~(b)

\noindent
is exothermic by 51.9 kJ/mol and shows an energy barrier of 24.3 kJ/mol. These results are in reasonable agreement with previous lower level calculations (Bouchoux \& Choret 1997; Fridgen et al. 2001). Alternatively, MIN1 can isomerize to the more stable species MIN2, overcoming a very small barrier of 4.6 kJ/mol. However, MIN2 cannot evolve to any other species, while MIN3 can isomerizes to MIN4 or MIN5 overcoming very small barriers (see from Fig. \ref{fig:reac1-pes} and Table \ref{tab:enthalpy}). MIN3, MIN4 and MIN5 are essentially electrostatic complexes of protonated dimethyl ether and water with a different geometry, as we can appreciate from the optimized geometries reported in Fig. \ref{fig:reac1-geometry}. These species dissociate to protonated dimethyl ether and water in endothermic reactions as we can see in Fig. \ref{fig:reac1-pes} and Table \ref{tab:enthalpy}, but globally the reaction is exothermic by 62.8 kJ/mol (in agreement with the experimental determination by Fridgen et al. 2001).

\begin{table*}
  \begin{tabular}{lcccc}
    \hline
    & \multicolumn{2}{c}{$\Delta$H$^0_0$} & \multicolumn{2}{c}{Barrier height} \\
    & B3LYP & CCSD(T)                                & B3LYP & CCSD(T) \\
         \hline
    CH$_3$OH + CH$_3$OH$_2^+$~$\rightarrow$~MIN1 & -40.2 & -47.3 & & \\
    MIN1~$\rightarrow$~MIN2 & -92.7 & -92.4 & 2.5 & 4.6 \\
    MIN1~$\rightarrow$~MIN3 & -48.1 & -51.9 & 16.4 & 24.3 \\
    MIN3~$\rightarrow$~MIN4 & -65.3 & -61.9 & 0.8 & 2.1 \\
    MIN3~$\rightarrow$~MIN5 & -5.8   & -6.2   & 1.3 & 2.5 \\
    MIN3~$\rightarrow$~CH$_3$OHCH$_3^+$ + H$_2$O & 31.0 & 36.4 & & \\
    MIN4~$\rightarrow$~CH$_3$OHCH$_3^+$ + H$_2$O & 95.6 & 98.3 & & \\
    MIN5~$\rightarrow$~CH$_3$OHCH$_3^+$ + H$_2$O & 36.8 & 42.6 & & \\
    CH$_3$OHCH$_3^+$ + NH$_3$~$\rightarrow$~MIN6 & -153.6 & -157.3 & & \\
    MIN6~$\rightarrow$~CH$_3$OHCH$_3$ + NH$_4^+$ & 95.4 & 99.1 & & \\
         \hline
  \end{tabular}
  \caption{Enthalpy changes and barrier heights (kJ/mol, 0 K) computed at the B3LYP/aug-cc-pVTZ and CCSD(T)/aug-cc-pVTZ levels of theory for selected reactions of the system CH$_3$OH + CH$_3$OH$_2^+$.}
\label{tab:enthalpy}
\end{table*}

All the involved intermediates and transition states lie under the energy of the reactants asymptote; reaction (1) should be, therefore, an efficient way to produce protonated dimethyl ether.
Regarding the energy values for each of the species participating in the CH$_3$OH + CH$_3$OH$_2^+$ reaction, we think that a comparison with the previous values calculated by Bouchoux \& Choret (1997), as well as by Fridgen et al. (2001), would be useful. With respect to reactants, our energy value for the initial O--C bound adduct is -47.3 kJ/mol compared to -56 kJ/mol obtained by Bouchoux \& Choret and -45 kJ/mol by Fridgen et al. Our value is between the two but much closer to the second one. On the other hand, regarding the TS between the initial and the protonated ether-water intermediate, our value is -23.0 kJ/mol, again between the two previous values (-15 kJ/mol by Fridgen et al., -26 kJ/mol by Bouchoux \& Choret). It is interesting that the barrier for the first rearrangement of the initial adduct is fortuitously the same for both previous authors (30 kJ/mol) while it is much lower in our case (24 kJ/mol). Regarding the second intermediate (where the water molecule is about to exit), our value (-99.2 kJ/mol) is once more between the two previous ones (-92 kJ/mol by Fridgen et al., -102 kJ/mol by Bouchoux \& Choret). This time the proportion of the three energy values is the same as in the case of the transition state and, hence, the barrier for back-rearrangement is essentially the same in all three cases. Finally, as regards the products energy, we predict a somewhat higher exothermicity (-62.8 kJ/mol) than those obtained by previous authors (-53 kJ/mol by Fridgen et al., -52 kJ/mol by Bouchoux \& Choret).

\subsubsection{The reaction (CH$_3$)$_2$OH$^+$ + NH$_3$}
\begin{figure*}
  \centering
  \includegraphics[width=17.0cm]{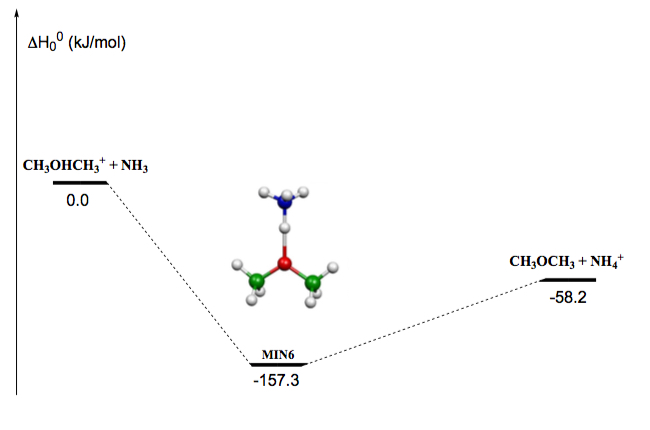}
  \caption{Schematic representation of the potential energy surface relative to the interaction between (CH$_3$)$_2$OH$^+$ and NH$_3$. For simplicity, only the CCSD(T) relative energies (kJ/mol) are reported. }
  \label{fig:reac2-pes}
\end{figure*}

The schematic PES of reaction (2) is reported in Fig. \ref{fig:reac2-pes} and the optimised geometry of reactants, intermediates and products is reported in Figs. \ref{fig:geometry} and \ref{fig:reac1-geometry}. This reaction is exothermic by 58.2 kJ/mol both at B3LYP/aug-cc-pVTZ and CCSD(T)/aug-cc-pVTZ level of calculations. The interaction between protonated dimethyl ether and ammonia gives rise, in a barrierless process, to an adduct (MIN6) more stable than the reactants by 153.6 kJ/mol at B3LYP/aug-cc-pVTZ level and by 157.3 kJ/mol at CCSD(T)/aug-cc-pVTZ level. The dissociation of this species into dimethyl ether and ammonium ion is endothermic by 95.4 kJ/mol at B3LYP/aug-cc-pVTZ level and by 99.1 kJ/mol at CCSD(T)/aug-cc-pVTZ level.

\subsection{Kinetics Calculations: Methods}\label{sec:deta-kinet-calc}
The kinetics of the two reaction schemes were investigated using capture theory and Rice-Ramsperger-Kassel-Marcus (RRKM) scheme, as used in other cases before (Skouteris et al. 2015, 2017, 2018; Sleiman et al. 2018; Balucani et al. 2018). In particular, the rate coefficient of each unimolecular step at a specific energy E is given by the expression:
\begin{equation}
k(E) = \frac{N(E)}{h~\rho(E)}
\end{equation}
where $N(E)$ ) represents the sum of states of the transition state at an energy E, $\rho$(E) represents the reactant density of states and $h$ is Planck's constant. The initial bimolecular association step is treated using capture theory, assuming that the entrance potential is of the form:
\begin{equation}
V(R) = - \frac{C_4}{R^4}
\end{equation}
(R being the distance between the two particles) as appropriate for a charge-dipole interaction. The rate of the inverse capture step (back-dissociation) is calculated from the capture rate constant and the densities of states of the reactants and the initial adduct, using a detailed balance argument. Finally, when no clear transition state is 
present (as in the final step of the protonated dimethyl ether formation) we use variational transition state theory, whereby the minimum rate constant is chosen among several ``candidate'' transition states along the reaction coordinate. Having obtained rate constants for all intermediate steps (at a specific energy), we make a steady state assumption for all intermediates and thereby, resolving a master equation, derive energy-dependent rate constants for the overall reaction (from initial reactants to final products). Finally, we do a Boltzmann averaging of the energy dependent rate constants to derive canonical rate constants (as a function of temperature). The rate constants as a function of temperature have been fitted to the modified Arrhenius form: 
\begin{equation}
k(T) = A ( \frac{T}{300} )^\beta exp^{-\frac{\gamma}{T}}
\end{equation}
for temperatures up to 600 K.

\subsection{Kinetics Calculations: Results}\label{sec:results-kinet-calc}

\subsubsection{The reaction CH$_3$OH + CH$_3$OH$_2^+$}
According to our calculations, the initially step of this reaction features the association of the two reactants with the formation of an O--C adduct (MIN1). Note that the formation of a proton bound dimer is not the favored initiating step, as instead suggested by Bouchoux \& Choret (1997) but disproved by Fridgen et al. (2001) on the basis of their experimental determination. Hence, MIN1 is formed directly from the reactants rather than from a rearrangement of the proton bound adduct. The O--C adduct has the option of undergoing back-dissociation to reactants or else rearranging to a second intermediate (MIN3), whereby the newly formed O--C bond is shortened and acquires covalent character, whereas the old O--C bond is weakened. Alternatively, MIN1 can rearrange to the intermediate MIN2 where two CH$_3$OH molecules are held together by a proton between them. However, the only energetically feasible option for this intermediate is to rearrange back to MIN1, and thus it makes no difference to the rate constants. In our path, the final step is the departure of an H$_2$O molecules, that terminates the reactive process. The H$_2$O molecule can depart from the MIN3 intermediate in more than one way. One path is direct, monotonic departure of H$_2$O. Bouchoux \& Choret identified the possibility of an internal rotation of the MIN3 intermediate, forming a new intermediate (MIN4) which is a proton-bound adduct between protonated dimethyl ether and H$_2$O. We have also found and included this path in our scheme, and it constitutes the second way of eliminating an H$_2$O molecule. Finally, MIN3 can rearrange to the dimer water - protonated dimethyl ether (MIN5 in our scheme) which can subsequently dissociate to products.

The rate coefficients were found to decrease monotonically with temperature. This, as found in other cases before, is an effect of the rate of back-dissociation of the initial adduct increasing with temperature much more rapidly than the rate of dimethyl ether formation. Because of the characteristic trend of the reaction, we found much more satisfactory fits separating the temperature range into three: a ``low'' temperature range, spanning temperatures from 10 to 50 K, a ``medium'' temperature range spanning 51--180 K and a ``high'' temperature range, spanning temperatures from 180 K to 600 K. Moreover, we saw that the fit is highly insensitive to the values of the $\gamma$ parameter and it is possible to obtain equally good fits for a wide range of its values. Therefore, we have chosen to set its value to 0 (as currently assumed in the KIDA and UMIST databases). Given these constraints, the optimal values of $\alpha$ and $\beta$ for the three temperature ranges turn out to be those reported in Table 2. The comparison of our results with the most recent and accurate experimental data of Fridgen et al. (between 293 and 338 K) confirms the accuracy of the present theoretical determination, as also shown in Fig. \ref{fig:rates}.

\begin{table}
  \begin{tabular}{ccc}
    \hline
    Temperature range (K) & $\alpha$ (cm$^3$s$^{-1}$) & $\beta$ \\
       \hline
    10--50     & $7.31\times10^{-10}$ & -0.113 \\
    51--180   & $1.58\times10^{-10}$ & -1.044 \\
    181--600 & $8.84\times10^{-11}$ & -2.075 \\
       \hline
  \end{tabular}
  \caption{Values of the $\alpha$ and $\beta$ coefficients, following the formalism in the KIDA database, of the reaction (1), namely CH$_3$OH + CH$_3$OH$_2^+$~ $\rightarrow$ ~ (CH$_3$)$_2$OH$^+$ + H$_2$O. Note that the rate coefficients refer to the three temperature ranges, as described in the text.}

\end{table}

For all the investigated temperatures, indeed, the ratio of the two (our and Fridgen et al.) values varies between 0.85--1.18. Both sets of rate conefficients diminish with a very similar temperature trend. For instance, at the lowest temperature investigated of 293 K we predict a rate coefficient of 9.5$\times10^{-11}$ cm$^3$s$^{-1}$, which compares well with the values by Fridgen et al. of 11.1$\pm 0.1 \times10^{-11}$ cm$^3$s$^{-1}$. At the highest temperature (338 K) the corresponding values are 6.9$\times10^{-11}$ cm$^3$s$^{-1}$ and 6.0$\pm 0.3 \times10^{-11}$ cm$^3$s$^{-1}$, respectively. 
Comparing our rate constants with those of Morris et al. (1991) at 300 and 450 K (and at the lowest pressures used by the authors, 0.26 Torr and 0.31 Torr respectively, in order to approach interstellar conditions as nearly as possible) we get 9.0$\times10^{-11}$ cm$^3$s$^{-1}$ at 300 K -compared to 7.0$\times10^{-11}$ cm$^3$s$^{-1}$ and  3.8$\times10^{-11}$ cm$^3$s$^{-1}$ at 450 K (compared to 4.2$\times10^{-11}$ cm$^3$s$^{-1}$). Finally, we mention the room temperature result of Dang \& Bierbaum (1992) of 6.1$\times10^{-11}$ cm$^3$s$^{-1}$ (applying the corrected branching ratio of the authors), again in reasonable agreement with our results.
On the contrary, the data by Karpas \& Meot-Ner (1989) are systematically higher than our values, but this is true also for all the other experimental results, possibly suggesting that their $k(T)$ values are too large.

\subsubsection{The reaction (CH$_3$)$_2$OH$^+$ + NH$_3$}
Regarding reaction (2), back-dissociation was found to be negligible. As a result, rate coefficients were found to be essentially independent of temperature, with constants $\alpha$ = 9.67$\times10^{-10}$ cm$^3$s$^{-1}$, $\beta$ = 0 and $\gamma$ = 0. There are no other experimental or theoretical data for this system to compare with our results. We can only note that our value is in agreement with the trend expected after the Hammond's postulate, being close to the lower limit of the values recorded by Hemsworth et al. (1974), that is ca. 1$\times10^{-9}$ cm$^3$s$^{-1}$.

% % % %
% % % %
% % % % 
%%%%%%%%%%%%%%%%%%%%%%%%%%%%%%%%%%%%%%%%%%%%%%%%%%%%%%%
\section{Discussion}\label{sec:discussion}
Figure 1 shows the rate coefficient of reaction (1) as computed in this work and compared with the available experimental data, namely those measured by Karpas \& Meot-Ner (1989), Morris et al. (1991), Fridgen et al. (2001) and Dang and Bierbaum (1992). The agreement between our calculations and the experimental data is very good, especially when considering the most recent set of data by Fridgen et al. (2001). This sustains the validity of the approach employed here. As we have already commented on, all experiments were carried out at room temperature or higher, so that we can state that our values are, at present, the best available rate coefficients of reaction (1) at the temperatures valid for the ISM. We emphasize again, therefore, that quantum chemistry calculations are, in some cases, the best, if not the only way to evaluate the rate of a reaction relevant to astrochemistry. This is even more evident in the case of reaction (2) for which no laboratory experiments exists. Our estimate provide a constant value of 9.67$\times10^{-10}$ cm$^3$s$^{-1}$ for all the temperatures of interest.

As already mentioned, the two major astrochemical databases, KIDA and UMIST, are the repository of the reactions rate coefficients used by different modelers. Therefore, a comparison with those value is in order. To this end, Fig. 1 compares the rate coefficients of reaction (1) computed in this work with those listed in KIDA and UMIST (namely the value also used by Taquet et al. 2016), respectively, as a function of the temperature. The agreement among the three estimates is relatively good, within a factor of two, up to about 100 K, with the old UMIST, namely the Taquet et al. (2016), value closer to ours. However, at lower temperatures the values diverge and, at 10 K, the Taquet et al. (2016) value overestimates the rate coefficient by more than a factor 20 with respect to our computed value.  The second step proposed by Taquet et al. (2016), reaction (2), is instead absent in both KIDA and UMIST.

In their work, Taquet et al. (2016) predicted that dimethyl ether is abundantly formed by reactions (1) and (2) in a gaseous environment with a temperature substantially at about 100 K. For their predictions, these authors used the value of reaction (1) in the old UMIST database and assumed a constant value of $\alpha$ = 2$\times10^{-9}$ cm$^3$s$^{-1}$ for reaction (2). Therefore, in the Taquet et al. work, while the rate of reaction (1) is underestimated by less than 20\%, the rate of reaction (2) is overestimated by about a factor two. We conclude that, within this factor two, the predictions by Taquet et al. are substantially correct, within the used model and adopted assumptions (e.g. temperature and ammonia abundance).

However, at very low temperatures (below 40 K), the use of both KIDA and UMIST rate coefficients is not warranted and we recommend the use of the present determination.

% % % %
% % % %
% % % % 
%%%%%%%%%%%%%%%%%%%%%%%%%%%%%%%%%%%%%%%%%%%%%%%%%%%%%%%
\section{Conclusion}\label{sec:conclusion}

In this work we have employed electronic structure and kinetics calculations to derive the reaction rate coefficients for two ion-molecule reactions recently proposed as a gas-phase formation route of dimethyl ether in interstellar objects. For reaction (1), the present calculations reproduce the scattered experimental results at high temperatures. In particular, they well reproduce the most recent and accurate data by Fridgen et al. (2001) and allow a more reliable extrapolation at the low temperatures of interest in interstellar objects (10-100 K). For reaction (2), the present calculations have allowed us to derive for the first time the value of the rate coefficient. This value is a factor 2 smaller than the one previously inferred by referring to similar processes.  The rate coefficients derived in this work essentially confirm the prediction by Taquet et al. (2016) concerning dimethyl ether formation in hot cores/corinos. Nevertheless, this formation route cannot  be efficient in cold objects (like prestellar cores) where dimethyl ether is also detected, because ammonia has a very low abundance in those environements.

% % % %
% % % %
% % % % 
%%%%%%%%%%%%%%%%%%%%%%%%%%%%%%%%%%%%%%%%%%%%%%%%%%%%%%%
\section*{Acknowledgments}
This work has been supported by MIUR PRIN 2015 funds,
project STARS in the CAOS (Simulation Tools for Astrochemical
Reactivity and Spectroscopy in the Cyberinfrastructure for Astrochemical
Organic Species), Grant Number 2015F59J3R. DS acknowledges
funding from SNS-Pisa Fondo Ricerca di Base. This work has been
supported by the project PRIN-INAF 2016 The Cradle of Life -
GENESIS-SKA (General Conditions in Early Planetary Systems for
the rise of life with SKA).
This project has received funding from the European Research Council (ERC) under the European Union's Horizon 2020 research and innovation programme, for the Project "The Dawn of Organic Chemistry" (DOC), grant agreement No 741002.

% % % %
% % % %
% % % % 
%%%%%%%%%%%%%%%%%%%%%%%%%%%%%%%%%%%%%%%%%%%%%%%%%%%%%%%
%\begin{thebibliography}{}
\section*{REFERENCES}

\noindent
Ag\'undez M. \& Wakelam V., 2013, Che.Rev., 113, 8710

\noindent
Balucani N., Skouteris D., Ceccarelli C., et al., 2018, Mol. Astrophys., 13, 30

\noindent
Balucani N., Skouteris D., Leonori F., et al., 2012, J. Phys. Chem. A, 116, 10467

\noindent
Balucani N., Ceccarelli C., Taquet V., 2015, MNRAS, 449, L16

\noindent
Barone V., Latouche C., Skouteris D., et al., 2015, MNRAS, 453, L31

\noindent
Bartlett R. J., 1981, Annu. Rev. Phys. Chem., 32, 359

\noindent
Becke A. D., 1993, J. Chem. Phys., 98, 5648

\noindent
Bouchoux G. \& Choret N., 1997, Rapid Commun. Mass Spectrom., 11, 1799

\noindent
Ceccarelli C., Caselli P., Fontani F. et al., 2017, ApJ, 850, 176

\noindent
Charnley S.B., Tielens A.G.G.M., Millar T.J., 1992, ApJ, 399, L71

\noindent
Charnley S.B., Kress M.E., Tielens A.G.G.M., Millar T.J., 1995, ApJ, 448, 232

\noindent
Codella, C., Ceccarelli, C., Caselli, P., et al. 2017, A\&A, 605, L3

\noindent
Dang T.T. \& Bierbaum V.M., 1992, Int. J. Mass Spectrom. Ion Processes, 117, 65

\noindent
Dunning T. H., Jr., 1989, J. Chem. Phys., 90, 1007

\noindent
Enrique-Romero, J., Rimola, A., Ceccarelli, C., Balucani, N. 2016, MNRAS, 459, L6

\noindent
Fridgen T.D., Keller J.D., \& McMahon T.B., 2001, J. Phys. Chem. A, 105, 3816

\noindent
Frisch M. J., Trucks G. W., Schlegel H. B.,  Scuseria G. E., Robb M. A., et al., Gaussian 09, Revision A.02, Gaussian, Inc., Wallingford CT (2009)

\noindent
Fl\"{u}kiger P.,  L\"{u}thi H. P., Portmann S., Weber J., MOLEKEL 4.3, Swiss Center for Scientific Computing, Manno (Switzerland), 2000

\noindent
Garrod, R. T. \& Herbst, E. 2006, A\&A, 457, 927

\noindent
Georgievskii Y., \& Klippenstein S., 2007, JPCA, 111, 3802

\noindent
Geppert W. D. et al., 2007, in European Planetary Science Congress 2007,
p. 613, available at: http://meetings.copernicus.org/epsc2007

\noindent
Gonzalez C., Schlegel H. B., 1989, J. Chem. Phys., 90, 2154

\noindent
Gonzalez C., Schlegel H. B., 1990, J. Phys. Chem., 94, 5523

\noindent
Hamberg M., \"{O}sterdahl F., Thomas R.D et al., 2010, A\&A 514, A83

\noindent
Herbst, E. \& van Dishoeck, E. F. 2009, ARA\&A, 47, 427

\noindent
Karpas Z. \& Meot-Ner (Mautner) M., 1989, J. Phys. Chem., 93, 1859

\noindent
Kendall R. A., Dunning  T. H. Jr., Harrison R. J., 1992,  J. Chem. Phys., 96, 6796

\noindent
Lamberts T., 2018, A\&A, 615, L2

\noindent
Leonori F., Skouteris D., Petrucci R., et al., 2013, J. Chem. Phys., 138, 024311

\noindent
Linnartz H., Ioppolo S., Fedoseev G., 2015, Int. Rev. Phys. Chem., 2015, 34, 205

\noindent
McElroy, D., Walsh, C., Markwick, A. J., et al. 2013, A\&A, 550, A36

\noindent
Morris R.A., Viggiano A.A., Paulson J.F., Henchman M.J., 1991, J. Am. Chem. Soc., 113, 5932

\noindent
Olsen J., Jorgensen P., Koch H. et al., 1996, J. Chem. Phys. 104, 8007

\noindent
Portmann S., L\"{u}thi H. P., 2000, Chimia 54, 766

\noindent
Qu\'enard D., Jim\'enez-Serra I., Viti S., Holdship J., Coutens A., 2018, MNRAS, 474, 2796

\noindent
Potapov A., Canosa A., Jim\'enez E., Rowe B., 2017, Angew. Chem., Int. Ed., 56, 8618

\noindent
Raghavachari K., Trucks G. W., Pople J. A. et al., 1989, Chem. Phys Lett., 157, 479

\noindent
Rimola A., Taquet V., Ugliengo P., Balucani N., Ceccarelli C., 2014, A\&A, 572, 70

\noindent
Rimola A.,  Skouteris D., Balucani N., Ceccarelli C.,Enrique-Romero J. et al., 2018, ACS Earth and Space Chem., 

\noindent
Rodgers S.D. \& Charnley S.B., 2001, ApJ, 546, 324

\noindent
Rodgers S.D. \& Charnley S.B. 2001, MNRAS, 320, L61

\noindent
Rosi M., Mancini L., Skouteris D. et al., 2018, CPL, 695, 87

\noindent
Ruaud M., Wakelam V., Hersant F., 2016, MNRAS, 459, 3756

\noindent
Skouteris D., Balucani N., Faginas-Lago N. et al., 2015, A\&A, 584, A76

\noindent
Skouteris D., Balucani N., Ceccarelli C. et al., 2018, ApJ, 854, 135

\noindent
Skouteris D., Vazart F., Ceccarelli C. et al., 2017, MNRAS, 468, L1

\noindent
Sleiman C., El Dib G., Rosi M. et al., 2018, Phys. Chem. Chem. Phys., 20, 5478

\noindent
Song L., \& K\"{a}stner ,J. 2016, J. Phys. Chem. Chem. Phys., 18, 29278

\noindent
Song L., \& K\"{a}stner ,J. 2017,  ApJ, 850, 118

\noindent
Stephens P. J., Devlin F. J., Chablowski et al., 1994, J. Phys. Chem.,  98, 11623

\noindent
Taquet V., Ceccarelli C., Kahane C., 2012, A\&A, 538, A42

\noindent
Taquet V., Wirstr\"{o}m E.S., Charnley S.B., 2016, ApJ, 821, 46

\noindent
Tedder J.M. \& Walker G.S., 1991, J. Chem. Soc. Perkin Trans., 2, 317

\noindent
Vasyunin A. I., Caselli P., Dulieu F., Jim\'enez-Serra I., 2017, ApJ, 842, 33

\noindent
Vazart F., Latouche, C. Skouteris D., et al., 2015, ApJ, 810, 111 (2015)

\noindent
Vazart F., Calderini D., Puzzarini C. et al., 2016, J. Chem. Theory Comput., 12, 5385

\noindent
Wakelam V., Herbst E., Loison J.-C. et al., 2012, ApJS, 199, 21

\noindent
Woon D. E.  \& Dunning T. H. Jr., J. Chem. Phys.,1993, 98, 1358

%\end{thebibliography}

%
%%%%%%%%%%%%%%%%%%%%%%%%%%%%%%%%%%%%%%%%%%%%%%%%%%
%
% Don't change these lines\par
\bsp %typesetting comment\par

\label{lastpage}

\end{document}